\documentclass[twocolumn,pra,aps,showpacs]{revtex4}

\usepackage{amsmath}
\usepackage{amssymb}
\usepackage[dvips]{graphicx}
\usepackage{pstricks}
\usepackage{pst-node}
\usepackage{pst-plot}

\newcommand{\qed}{\hspace*{\fill}$\square$}

 \newcommand{\R}{\mathbf{R}}

 \newcommand{\Z}{\mathbf{Z}}

 \newcommand{\paratodo}{\forall\,}
 
 \newcommand{\vect}[1]{\boldsymbol{\mathrm{#1}}}
 
 \newcommand{\sset}[1]{ \{#1\} }

 \newcommand{\ket}[1]{|#1\rangle}

 \newcommand{\peso}[1]{\mathrm{wt}(#1)}

\begin{document}

\title[Short Title]{
Topological Computation without Braiding}

\author{H. Bombin and M.A. Martin-Delgado}
\affiliation{ Departamento de F\'{\i}sica Te\'orica I, Universidad
Complutense, 28040. Madrid, Spain. }

\begin{abstract}
We show that universal quantum computation can be performed within
the ground state of a topologically ordered quantum system, which
is a naturally protected quantum memory. In particular, we show
how this can be achieved using brane-net condensates in 3-colexes.
The universal set of gates is implemented without selective addressing
of physical qubits and, being fully topologically protected, it
does not rely on quasiparticle excitations or their braiding.

\end{abstract}


\maketitle

Topological quantum computation offers the possibility of
implementing a fault-tolerant quantum computer avoiding the
extremely low threshold error rates found with the standard quantum
circuit model \cite{kitaev97}, \cite{freedman98},
\cite{freedman_etal01}, \cite{preskillnotes}. Physical systems
exhibiting a topological quantum ordered state \cite{wenniu90},
\cite{wenbook04} can be used as naturally protected quantum memories
\cite{kitaev97}, \cite{dennis_etal02}, \cite{bravyikitaev98}.
Characteristic properties of topologically ordered systems are the
energy gap between ground state and excited states,
topology-dependent ground state degeneracies, braiding statistics
of quasi-particles, edge states, etc. \cite{wenbook04}. The idea
is then to place the information in the topologically degenerate
ground state of such a system, so that the protection of the
encoded information comes from the gap and the fact that local
perturbations cannot couple ground states. In fact, the
probability of tunneling between orthogonal ground states is
exponentially suppressed by the system size and vanishes in the
thermodynamic limit.

A stabilizer code \cite{gottesman96}, \cite{calderbank_etal97} can
be topological. The best known example are Kitaev's surface codes
\cite{kitaev97}, \cite{dennis_etal02}. In general a code is
topological if its stabilizer has local generators and
non-detectable errors are topologically non-trivial (in the
particular space where the qubits are to be placed). Given such a
code, one can always construct a local Hamiltonian such that the
resulting system is topologically ordered and the error correcting
code corresponds to the ground state. An explicit example of this
Hamiltonian construction is given later on eq. \eqref{Hamiltoniano}.
Errors in the code amount to excitations.

Although the storage of quantum information is interesting by
itself, one would like to perform computations on it. A natural
approach in this context is that of considering a topological
stabilizer code in which certain operators can be implemented
transversally, which avoids error propagation within codes.
In terms of the corresponding topologically ordered system, this
means that operations are implemented without selective addressing
of the physical subsystems that make up each memory. This is
important for physical applications.

Unfortunately, surface codes only allow the transversal
implementation of the CNot gate.
Then the problem arises of wether there exists a topological
stabilizer code in which a universal set of gates can be performed
transversally. In fact, even at the level of general codes it is a
difficult task to find such codes \cite{klz96}. For most codes,
additional tricks such as the generation of large cat states are
unavoidable. However, quantum Reed-Muller codes have the very
special property of allowing the transversal implementation of the
gates:
\begin{equation}\label{reed-muller_gates}
K^{1/2} = \begin{pmatrix} 1 & 0 \\ 0 & i^{1/2}\end{pmatrix}, %
\Lambda = \begin{pmatrix} I_2 & 0 \\ 0 & X\end{pmatrix},
\end{equation}
where $X$ is the usual $\sigma_1$ Pauli matrix. Complemented with
the ability to initialize eigenstates of $X$ and $Z$ and to measure
this operators, these gates are enough to perform arbitrary
computations.
In particular, the Hadamard gate can be reconstructed and the set of
gates $\sset{H, K^{1/2}, \Lambda}$ is known to be universal
\cite{universal_raizK2}.

In this paper we will construct a 3-dimensional system showing
topological quantum order in which the gates
\eqref{reed-muller_gates} can be implemented. The ground state of
the system is a topological stabilizer code.
No other topological code of any dimension is known such that the
transversal implementation of a universal set of gates is possible.
In fact, a key ingredient in our approach is the appearance of
membranes\cite{hzw05}.
Our system is a
3-dimensional lattice with qubits located at the sites, and the
operations on the ground state are implemented without any selective
addressing of these physical qubits. This is in contrast with the
current approach to topological computation which relies on the
topological properties of quasiparticle excitations instead of
ground states properties and needs a selective braiding of
quasiparticles to produce quantum gates.
In fact our system is abelian, in the sense that monodromy
operations on excitations can give rise only to global phases.
In contrast,
in the context of quasiparticle braiding abelian systems can never
give universal computation. Therefore, our results enlarge the range
of applicability of the topological approach to quantum computation.

\begin{figure}
\includegraphics[width=5 cm]{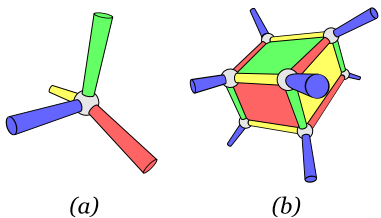}
\caption{(a) A generic site in a 3-colex.
 (b) The neighborhood of a particular b-cell:
faces are colored according to the color of their visible side (they
are br-, bg- and by-faces).} \label{figura_lattice}
\end{figure}

To achieve this goal, we start with a brief description of the
topologically ordered 3-dimensional condensed matter systems
\cite{topo3d} that we need for our construction. Consider a lattice
with coordination number 4 in which links are colored with four
colors as in Fig.~\ref{figura_lattice}(a). Color is introduced as a
bookeeping tool to keep track of the different sites, links, faces
and cells in the 3D lattice. We will use red, green, blue and yellow
labels (r,g,b,y) as colors. Assume that the cells can also be
colored, in such a way that, for example, the boundary links of a
red cell is a net with coordination number 3 formed by green, blue
and yellow links, as in Fig.~\ref{figura_lattice}(b). We call those
3D lattices with this set of properties \emph{3-colexes}.
For any color $q$, $q$-links connect $q$-cells. A face lying between
a r- and a y-cell has a boundary link made up of g- and b-links. We
call such a face a ry-face.

At each site of the lattice we place a qubit. We will be considering
operators of the form
\begin{equation}\label{operador_celda}
B_S^\sigma := \bigotimes_{i=1}^n \sigma^{f_i},\quad \sigma = X,Z,
\quad f_i = \begin{cases}0 & i \not\in S, \\ 1 & i\in S \end{cases},
\end{equation}
where $S$ is a given set of qubits in the system, $n$ the total
number of qubits. The Hamiltonian proposed in \cite{topo3d} is
\begin{equation}\label{Hamiltoniano}
H= - \sum_{c\in C} B_c^X - \sum_{f\in F} B_f^Z,
\end{equation}
where $C$ and $F$ are the cells and faces of the lattice,
respectively. It gives rise to topological order. In particular, the
degeneracy of the ground state is $2^k$ with $k=3h_1$, where $h_1$
is the number of independent cycles of the 3-manifold in which the
lattice is built. In particular, in a 3-sphere $h_1=0$ and there is
no degeneracy at all, whereas in a 3-torus $h_1=3$. In topology,
$h_1$ is known as a Betti number \cite{nakahara2003}.

The ground states $\ket \psi$ of \eqref{Hamiltoniano} are
characterized by the conditions
\begin{alignat}2
 \paratodo c\in C\quad\quad &B_c^X \ket
\psi &= \ket \psi, \label{condiciones_cell}\\
\paratodo f \in F\quad\quad &B_f^Z \ket
\psi &= \ket \psi. \label{condiciones_face}
\end{alignat}
In fact cell and face operators commute, and the ground state is a
stabilizer quantum error correcting code
\cite{gottesman96},
 \cite{calderbank_etal}, \cite{kitaev97}. Those eigenstates $\ket
{\psi'}$ for which any of the conditions \eqref{condiciones_cell},
\eqref{condiciones_face} are violated is an excited state or, in
code terms, an erroneous state.

\begin{figure}
\includegraphics[width=6 cm]{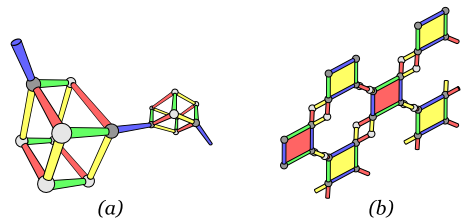}
\caption{(a) A b-string consists of several b-links that connect
b-cells.
(b) A ry-membrane is made up of ry-faces linked by bg-faces.
bg-faces are not shown here, only their links.
}\label{figura_string_membrane}
\end{figure}

Both excitations and degeneracy are best understood introducing
string and membrane operators.
A $q$-string, for some color $q \in \{ r,g,b,y \}$, is a collection
of $q$-links, as in Fig.~\ref{figura_string_membrane}(a). Strings
can have endpoints, which are always located at $q$-cells. Along
with every $q$-string $s$ we introduce the string operator $B_s^Z$.
If $\ket \psi$ is a ground state, then $B_s^Z \ket \psi$ is in
general an excited state, with excitations or quasiparticles at
those $q$-cells that are endpoints of $s$. If $s$ is closed, that is,
if it has no endpoints, $B_s^Z$ commutes with the Hamiltonian
\eqref{Hamiltoniano}.

Similarly, a collection of $pq$-faces, for any colors $p$ and $q$,
is a $pq$-menbrane, as in Fig.~\ref{figura_string_membrane}(b). For
any $pq$-membrane $m$ the corresponding membrane operator is
$B_m^X$. If $\ket \psi$ is a ground state and $m$ a rg-membrane, for
example, then $B_m^X \ket \psi$ is in general an excited state, with
excitations at those by-faces that form the border of $m$. These
excitations are closed fluxes  crossing the excited faces.
As an example, consider a ry-membrane such as the one in
Fig.~\ref{figura_string_membrane}(b). Its border will create a
ry-flux, which will cross those bg-faces at the border of the
membrane.
If $m$ is closed, that is if it has no borders, then $B_m^X$
commutes with the Hamiltonian \eqref{Hamiltoniano}.

As long as we consider closed manifolds in 3D, closed strings and
membranes are enough to form a basis from which any operator that
leaves the ground state invariant can be constructed. There are
three key points here.
First, any two string or membrane operators which are equal up to a
deformation have the same action on the ground state, which is in
itself a uniform superposition generated by all the possible local
deformations.
Second, a $q$-string operator $B_s^Z$ and a $pq$-membrane operator
$B_m^X$ anticommute iff $s$ crosses $m$ an odd number of times.
Otherwise, they commute and the same is true if they do not share
any color. Third, not all colors are independent. For example, the
combination of a r-, a g- and a b-string gives a y-string. In fact ,
there are exactly 3 independent colors for strings and 3 independent
color combinations for membranes. Therefore, all that matters about
strings and membranes is their color and topology, and the
appearance of the number 3 in the degeneracy is directly related to
the number of independent colors.

\begin{figure}
\includegraphics[width=5.7 cm]{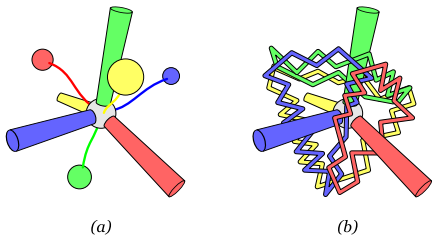}
\caption{(a) The $Z$ operator of a site creates one quasiparticle at
each of the cells that meet at the site. (b) The $X$ operator of a
site creates the flux structure shown, which corresponds to a flux
excitation at each of the faces meeting at the site.
}\label{figura_elementales}
\end{figure}

\begin{figure}
\includegraphics[width=6.5 cm]{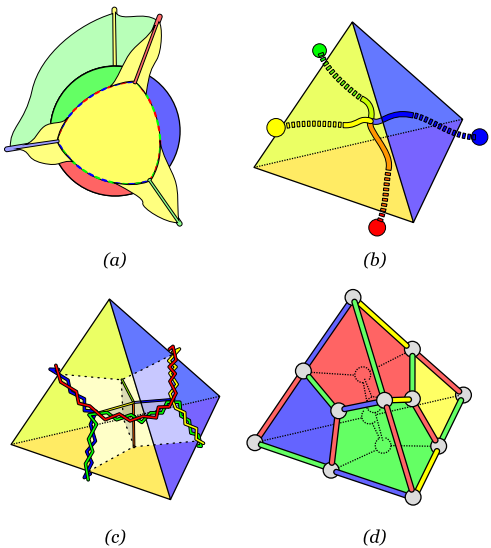}
\caption{(a) Here we represent the 3-sphere as $\R^3$ plus the point
at infinity  where we place the site to be removed. The faces and
links perpendicular to the colored sphere are partially displayed
but they continue to the infinity. These faces and links must be
removed as well. After their removal, we get a solid 2-sphere with a
surface divided in four triangular areas. This colored sphere
represents the remanining 3-colex itself. Then it can be reshaped to
get a tetrahedron. (b) A nontrivial string-net in the tetrahedron.
Its endpoints lye on the missing cells. (c) A nontrivial
membrane-net configuration in the tetrahedron. Its borders create
fluxes that cross the missing faces. Branching lines have been
suitably colored. (d) The simplest tetrahedral lattice. Here colors
have been given both to links and to cells. In the language of error
correction, it is a $[[15,1,5]]$ code, that is, it encodes a qubit
in 15 physical qubits, whereas its distance is 5 an so corrects up
to 2 errors.}\label{figura_tetrahedro}
\end{figure}

On the other hand, strings and membranes with a single color are not
enough to describe excitations. In general, strings can form a net
with branchpoints at which four strings meet, one for each color
(see Figs.\ref{figura_elementales}(a),\ref{figura_tetrahedro}(b))
Likewise, membranes can form nets in which, for example, a gb-, a
br- and a rg-membrane meet along a line (see
Fig.\ref{figura_tetrahedro}(c)). In order to study the exact
properties of general excitations, one can consider the elementary
excitations attached to the operators $X$ and $Z$ at any particular
site $i$ of the lattice. Let $\ket \psi$ be a ground state. Then the
state $Z_i\ket \psi$ is an excited state with four quasiparticles,
see Fig.~\ref{figura_elementales}(a). The state $X_i\ket \psi$ is an
excited state with six elementary fluxes which can be arranged in
four single color closed fluxes, as in
Fig.~\ref{figura_elementales}(b). From this class of elementary
excitations one can build any general excitation.

If we restrict ourselves to closed manifolds, there is no way in
which we can have a ground state with twofold degeneracy, or
equivalently, that encodes a single qubit. However, we will now
explain how one can obtain such a system by puncturing a 3-manifold.
In particular, consider any 3-colex in a 3-sphere. The ground state
in this case is non-degenerate. Now we choose any site in the
lattice and remove it. Moreover, we also remove the four links, six
faces and four cells that meet at the site. As a result, we obtain a
lattice with the topology of a solid 2-sphere, see
Fig.~\ref{figura_tetrahedro}(a).
 In order to calculate the degeneracy of the new system, we note that we have
removed one physical qubit and two independent generators
\cite{topo3d} of the stabilizer.
This is so because i/ although we remove 4 cells, three of the cell
operators can be obtained from the remaining one and the rest of
cell operators, see\cite{topo3d} and ii/ although we remove 6 faces,
5 of the face operators can be obtained from the remaining one and
the rest of face operators in the corresponding cells.
 Then, from
the theory of stabilizer codes it readily follows that the new code
encodes one qubit. This can also be understood using strings and
membranes. The surface of the system is divided into four faces,
each of them being the boundary with one of the removed cells. Thus,
we can color these areas with each color of the faces from the
removed cells, as in Fig.~\ref{figura_tetrahedro}(a). It is natural
to deform this sphere into a tetrahedron, and we will do so. Then
each of its faces can be the endpoint of a string of the same color,
and thus there is a single independent nontrivial configuration for
a string-net, as depicted in Fig.~\ref{figura_tetrahedro}(b). This
configuration, of course, corresponds to a string-net operator that
creates one quasiparticle excitation at each missing cell. In a
similar fashion, one can consider a net of membranes that creates
the flux configuration shown in Fig.~\ref{figura_tetrahedro}(c).
This net consists of six membranes, meeting in groups of three at
four lines that meet at a central point.
Observe that these excitations are in exact correspondence with
those in Fig.~\ref{figura_elementales}, when we see them from the
point of view of the removed site.

Although these string-net and membrane-net operators just discussed
can be used to introduce an operator basis for the encoded qubit,
this can be done in an alternative way that is more convenient for
practical implementations. Given any operator $O$ that acts on a
single qubit, we will use the notation $\hat {O}:= O^{\otimes n}$
for the operator that applies $O$ to each of the $n$ physical qubits
in the 3D lattice. Then, in any tetrahedral lattice we have $
\sset{\hat X,\hat Z} = 0$, because the total number of sites is odd:
every 3-colex has an even number of sites and we have removed one.
(see Fig.\ref{figura_tetrahedro}(d) for $n=15$).
 Since both $\hat X$ and $\hat Z$ commute with the
Hamiltonian \eqref{Hamiltoniano}, they can be considered the $X$ and
$Z$ Pauli operators on the protected qubit. As usual, let $\ket 0$
and $\ket 1$ be  a positive and a negative eigenvector of $Z$,
respectively, so that they form an orthogonal basis for the qubit
state space. Let also $\ket {\vect v} :=
\ket{v_1}\otimes\dots\otimes\ket{v_n}$ be a vector state for any
binary vector $\vect v\in \Z_2^n$, $\Z_2 = \sset{0,1}$. These binary
vectors are usually introduced in error correcting codes
\cite{calderbank_etal97}. A basis for the protected qubit can be
constructed as follows
\begin{align}\label{base_protegida}
 \ket {\hat 0} &:= \prod_c (1+B_c^X) \ket {\vect 0} = \sum_{\vect v\in V} \ket {\vect
 v},\\
 \ket {\hat 1} &:= \prod_c (1+B_c^X) \ket {\vect 1} = \hat X \ket {\hat 0} = \sum_{\vect v\in V} \ket {\bar {\vect v}},
\end{align}
where, $\vect 0 := (0\cdots 0)$, $\vect 1 := (1\cdots 1)$, $\bar
{\vect v}:= \vect 1+\vect v$, $c$ runs over all cells in the lattice
and $V$ is the subspace spanned by vectors ${\vect v}_c$ such that
$\ket {{\vect v}_c} = B_c^X\ket {\vect 0}$. In order to be able to
apply the $K^{1/2}$ gate to the protected qubit in the tetrahedral
lattice, we must introduce a new requirement. We impose that faces
(cells) must have a number of sites which is a multiple of four
(eight). The simplest example of such a tetrahedral lattice is
displayed in Fig.\ref{figura_tetrahedro}(d). As we will show below,
it follows from these conditions that
\begin{equation}\label{congruencia}
    \paratodo\vect v\in V \quad \peso {\vect v}\equiv 0\mod 8,
\end{equation}
where the weight of a vector $\peso {\vect v}$ is the number of 1s
it contains. But then we have
\begin{equation}\label{base_protegida}
 \hat K^{1/2}\ket {\hat 0} = \ket {\hat 0},\qquad
 \hat K^{1/2}\ket {\hat 1} = i^{l/2} \ket {\hat 1}
\end{equation}
where $l\equiv n \mod 8$, $l\in\sset{1,3,5,7}$. This means that the
global $\hat K^{1/2}$ operator can be used to implement $K^{1/2}$ on
the encoded qubit, by repeated application in the case that $l\neq
1$.

We still have to prove \eqref{congruencia}. Let the weight of a
Pauli operator be the number of sites on which it acts nontrivially,
and let us work modulo 8. Then \eqref{congruencia} says that for any
product $\pi = B_{c_1}^X\dots B_{c_m}^X$, $\peso \pi\equiv 0$. This
follows by induction on $m$. The case $m=0$ is trivial. For the
induction step, we first observe that if $\peso \pi\equiv 0$, then
$\peso{\pi B_c^X}\equiv 0$ iff $\pi$ and $B_c^X$ share $s$ sites
with $s\equiv 0,4$. But if $f_1,\dots,f_j$ are those faces of $c$
shared with some cell of $\pi$, then $s=\peso{B_{f_1}^Z\cdots
B_{f_j}^Z}$. These faces are part of the 2D color lattice that forms
the boundary of the cell $c$, from which it follows  that $s\equiv
0,4$ \cite{topoDistillation}.

The $\Lambda$ gate \eqref{reed-muller_gates}, known as the CNot
gate, is more straightforward. Imagine that we take two identical
tetrahedral lattices and superpose them so that corresponding sites
get very near. Then we apply $\hat \Lambda$, that is, we apply
$\Lambda$ pairwise. This can be achieved through single qubit operations
and Ising  interactions. As a result,  it is easily checked that we get a
$\Lambda$ gate between the protected qubits.

As for measurements, the situation is the same as in any CSS code
\cite{calderbankshor96}, \cite{steane96b}. If we measure each
physical qubit in the $Z$ basis, then we are also performing a
destructive measurement in the $\hat Z$ basis. Then non-destructive
measurements of $\hat Z$ can be carried out performing a CNot gate
with the qubit to be measured as source and a $\ket {\hat 0}$ state
as target, and measuring the target destructively. Similarly, if we
measure each physical qubit in the $X$ basis we are performing a
measurement in the $\hat X$ basis. We can admit faulty measurements,
since the faulty measurement of a qubit is equivalent to an error
prior to it. Thus the measuring process is as robust as the code
itself and is topologically protected \cite{dennis_etal02}. The
results of the measurements must be classically processed to remove
errors and recover the most probable codeword.

Initialization is always a subtle issue in quantum computation,
whether topological or not, and it certainly depends upon the
physical implementation. In any case, even if perfectly pure $\ket
{\hat 0}$ or $\ket {\hat +}$ states cannot be provided, one can
still purify them as much as necessary if their fidelity is above
$\frac 1 2$. To do this, only the CNot gate $\hat \Lambda$ and
measurements in the $\hat Z$ and $\hat X$ bases are necessary.

As a concluding remark, we observe that the lattice that we have
described so far unifies the strategies used in fault tolerant
computation, such as transversal operations, with the concept of a
topologically protected quantum memory. Note that this approach is
very different from the usual one in topological quantum
computation, based on the braiding of non-abelian anyons in a two
dimensional system. In fact, the topological order of the
3-dimensional system that we have described is Abelian.

\noindent {\em Acknowledgements} We acknowledge financial support
from  an EJ-GV fellowship (H.B.), DGS grant  under contract BFM
2003-05316-C02-01 and EU project INSTANS (M.A.MD.), and CAM-UCM
grant under ref. 910758.

\end{document}